\documentclass[epj]{webofc}
\usepackage[varg]{txfonts}   
\usepackage{multirow} 
\usepackage{eepic}
\usepackage[nooneline]{caption} 
\usepackage{sidecap} 
\usepackage{wrapfig}
\usepackage{floatrow}
\wocname{European Physical Journal Web of Conferences}
\woctitle{International Conference on New Frontiers in Physics 2013}
\begin{document}
\title{$Q_{\rm weak}$: First Direct Measurement of the Weak Charge of the Proton}
%
%

\author{Nuruzzaman\inst{1}\fnsep\thanks{\email{nur@jlab.org}} 
(for The $Q_{\rm weak}$ Collaboration)}

\institute{Hampton University, Hampton, VA, 23668, USA}

\abstract{%
The $Q_{\rm weak}$ experiment at Hall C of Thomas Jefferson National Accelerator Facility has made the first direct measurement of the weak charge of the proton, $Q^{p}_{W}$, through a precision measurement of the parity-violating asymmetry in elastic e-p scattering at low momentum transfer $Q^{2}$= 0.025 (GeV/c)$^{2}$ with incident electron beam energy of 1.155 GeV. The $Q_{\rm weak}$ experiment, along with earlier results of parity violating elastic scattering experiments, is expected to determine the most precise value of $Q^{p}_{W}$ which is suppressed in the Standard Model. 
If this result is further combined with the $^{133}$Cs atomic parity violation (APV) measurement, significant constraints on the weak charge of the up quark, down quark, and neutron can be extracted.
This data will also be used to determine the weak-mixing angle, $\sin^{2} \theta_{W}$, with a relative uncertainty of < 0.5\% that will provide a competitive measurement of the running of $\sin^{2} \theta_{W}$ to low $Q^{2}$. 
An overview of the experiment and its results using the commissioning dataset, constituting approximately 4\% of the data collected in the experiment, are reported here.
}
\maketitle
%
\section{Introduction}
\label{intro}



The Standard Model (SM) is the most successful elementary particle theory developed so far. However, it is considered to be a low energy effective theory due to the missing phenomena like gravity and dark matter. The weak charge of the proton, $Q^{p}_{W}$, is the neutral current analog to the proton's electric charge. The SM precisely predicts the weak charge of the proton (Table~\ref{tab-1}), which is suppressed in the SM. It is therefore a good candidate for an indirect search~\cite{ref-musolf1994, ref-Erler2003, ref-Marciano1985, ref-Erler2005, ref-Erler2005_2} for new parity-violating (PV) physics between electrons and light quarks.

The $Q_{\rm weak}$ experiment~\cite{qweak_proposal_2007} at Hall C of Thomas Jefferson National Accelerator Facility (TJNAF) has made the first direct measurement of the weak charge of the proton through the precision measurement of the PV asymmetry in elastic e-p scattering at low momentum transfer $Q^{2}$= 0.025 (GeV/c)$^{2}$ with incident electron beam energy of 1.155 GeV. The experiment was performed during the period of November 2010 to May 2012. Herein, we report the results obtained from only about 4\% of the total data collected in the experiment. 

\begin{table}[!h]
\centering
\begin{tabular}{c | c | c l c }
\hline
Particle & EM & & Weak Charge & \\ \hline
\small u & \small 2/3 & \small -2$C_{1u}$ & \small =1-(8/3)$\sin^{2}\theta_{W}$ & \small$\approx$1/3 \\
\small d & \small -1/3 & \small -2$C_{1d}$ & \small =-1+(4/3)$\sin^{2}\theta_{W}$ & \small$\approx$-2/3 \\
\small p(uud) & \small 1 & \small $Q^{p}_{W}$ -2$(2C_{1u}+C_{1d})$ & \small =1-4$\sin^{2}\theta_{W}$ & \small$\approx$0.07 \\
\small n(udd) & \small 0 & \small $Q^{n}_{W}$=-2$(C_{1u}+2C_{1d})$ &  & \small$\approx$-1 \\
\hline
\end{tabular}
\caption{The Standard Model predictions of the electromagnetic  and weak charges (from Z pole measurement) of quarks and nucleons. The weak charge of proton is suppressed in the SM.}
\label{tab-1}       
\end{table}



The measured asymmetry in the experiment is the difference over the sum of the elastic e-p scattering yield for electrons with positive and negative helicity and is expressed as,

\begin{equation} \label{equ:eqAsym}
A_{ep} = \frac{Y^{+} - Y^{-}}{Y^{+} + Y^{-}}
\end{equation}


At tree level this asymmetry can be described in terms of electromagnetic (EM) form factors~\cite{ref-Kelly2004} $G^{\gamma}_{E}$, $G^{\gamma}_{M}$, weak neutral form factors $G^{Z}_{E}$, $G^{Z}_{M}$, and the neutral weak axial form factor $G^{Z}_{A}$~\cite{ref-musolf1994, ref-Armstrong2012} as

\begin{equation} \label{equ:eqAsym1}
A_{ep} = \left[ \frac{-G_{F}Q^{2}}{4\pi\alpha\sqrt{2}} \right]\times \\
\left[ \frac{\epsilon G^{\gamma}_{E}G^{Z}_{E} + \tau G^{\gamma}_{M}G^{Z}_{M} - \left(1-4\sin^{2}\theta_{W} \right)\epsilon^{\prime}G^{\gamma}_{M}G^{Z}_{A}  }{\epsilon\left(G^{\gamma}_{E}\right)^{2} + \tau\left(G^{\gamma}_{M}\right)^{2}   } \right]
\end{equation}

where kinematic quantities can be written as 

\begin{equation} \label{equ:eqAsym2}
\epsilon = \frac{1}{1+2(1+\tau)\tan^{2}\frac{\theta}{2}}, \epsilon^{\prime} = \sqrt{\tau(1+\tau)(1-\epsilon^{2})}
\end{equation}

$G_{F}$ is the Fermi constant, $\sin^{2} \theta_{W}$ reflects the weak mixing angle, $Q^{2}$ is the negative four momentum transfer squared,  $\alpha$ is the fine structure constant, $\tau$ = $Q^{2}$/4$M^{2}$, M is the proton mass, and $\theta$ is the electron scattering angle in the laboratory frame. If we express the factor $\left[ \frac{-G_{F}Q^{2}}{4\pi\alpha\sqrt{2}} \right] = A_{0}$, then the reduced asymmetry in the forward angle limit ($\theta \rightarrow$ 0) and at low $Q^{2}$ can be written conveniently  \cite{ref-Yong2007} as

\begin{equation} \label{equ:eqAsym3}
A_{ep}/A_{0} = Q^{p}_{W} + Q^{2}B(Q^{2},\theta)
\end{equation}

Then $Q^{p}_{W}$ is the intercept of $A_{ep}$/$A_{0}$ \textit{vs.} $Q^{2}$ fit in eq.~\ref{equ:eqAsym3}. The term $Q^{2} B(\theta, Q^{2})$ is determined experimentally from existing parity violating electron scattering (PVES) data at higher $Q^{2}$ and is suppressed at low $Q^{2}$. It contains only the nucleon structure defined in terms of EM, strange quark, and weak form factors. The $Q^{2}$ of the measurement reported here is 4 times smaller than of any previously reported e-p PV experiment, which ensures a reliable extrapolation to $Q^{2} \rightarrow$ 0 using eq.~\ref{equ:eqAsym3}.

\begin{figure}[!h]
\centering
\includegraphics[width=14.0cm]{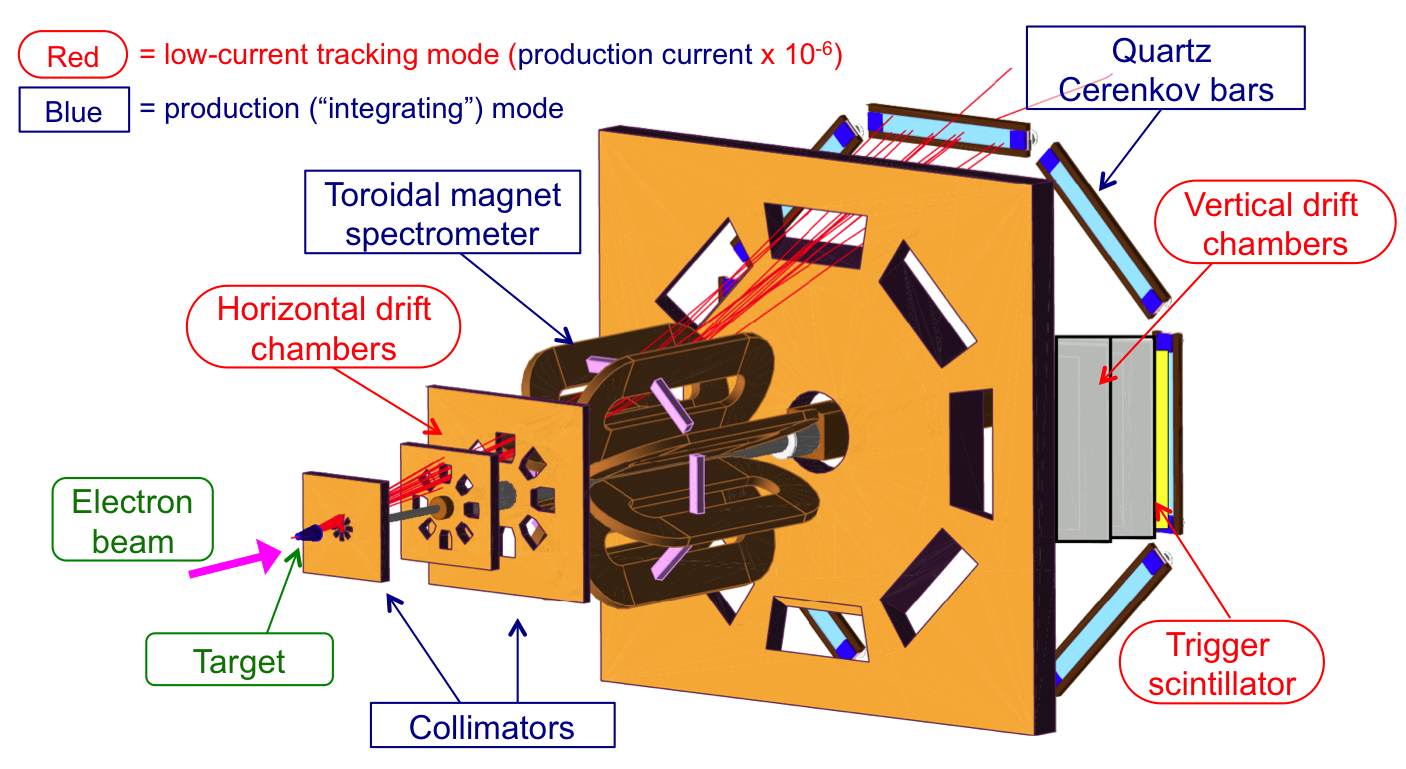}
\caption{(Color online) Schematic diagram of the $Q_{\rm weak}$ apparatus. The basic experimental design showing the target, collimators, toroidal magnet coils, electron trajectories, and detectors. Elastically scattered electrons focus at the \v{C}erenkov detectors. High current production mode apparatus components are shown in blue rectangular boxes and low current tracking mode components are shown in red elliptical boxes. Beam direction is from left to right. 
}
\label{figQweakApparatus}
\end{figure}

\section{Apparatus}
\label{Apparatus}

As the $Q_{\rm weak}$ PV asymmetry and its required absolute uncertainty are an order of magnitude smaller than in previous experiments, a dedicated design, significant improvement to hardware and software, and additional control of systematic uncertainties were needed to reach the proposed precision goals.

%


A longitudinally polarized electron beam was reversed at a rate of 960 Hz in a sequence of ‘‘pseudorandom helicity quartets’’ ($\pm\mp\mp\pm$), which helps minimize noise due to slow linear drifts, and limits noise due to fluctuations in the target density and in beam properties. 
The beam current was measured using radio frequency resonant cavities. Five beam position monitors (BPMs) upstream of the target were used to extract the beam position and angle at the target whereas energy changes were measured using a BPM at the point of highest dispersion in the beam line.
The electron beam of 180 $\mu$A scattered from a 34.4 cm long liquid hydrogen target \cite{qweakGreg} contained in an aluminum alloy container. A series of collimators were used to define the scattered electrons. A toroidal magnet focused elastically scattered electrons onto eight radiation-hard synthetic fused quartz \v{C}erenkov detectors which were symmetrically oriented about the beam axis at a radius of 3.4 m, 12.2 m downstream from the target (Figure~\ref{figQweakApparatus}). This azimuthal symmetry of the detectors helped reduce uncertainties from helicity-correlated beam motion and transverse beam polarization. Two photo multiplier tubes (PMTs) were optically coupled on the end of the detector bars. The PMT anode current fed custom low noise current to voltage preamplifiers whose output signals were digitized with 18 bit analog to digital converters (ADCs) sampling at 500 kHz.
Basic parameters and typical operating conditions during the experiment are described in Table~\ref{tab:qweak_parameters}.

\begin{table}[!h]
\begin{center}
  \begin{tabular}{ l  l }
    \hline
	Parameter & Value\\ \hline
	Incident beam energy & 1.155 GeV$\pm$0.003 GeV at target center\\
	Beam polarization & 89\%\\
	Beam current & 145 (180) $\mu$A\\
	LH$_{2}$ target thickness & 34.4 cm \\
	Cryopower & 2.5 kW\\
	Production running time & 2544 hours\\
	Nominal scattering angle & 7.9$^{o}$\\
	Scattering angle acceptance & $\pm$3$^{o}$\\
	Acceptance & 49\% of 2$\pi$\\
	Solid angle & $\Delta \Omega$ = 37 msr\\
	Acpt. avg. $Q^{2}$ & $\langle Q^{2}\rangle$ = 0.025 (GeV/c)$^{2}$\\
	Acpt. avg. physics asym. & $\langle A \rangle$ = -234 ppb\\
	Acpt. avg. experimental asym. & $\langle A \rangle$ = -200 ppb\\
	Luminosity & 2$\times$10$^{39}$ s$^{-1}$ cm$^{-2}$\\
	Integrated cross section & 4.0 $\mu$b\\
	Integrated rate & 6.5 GHz (0.81 GHz/sector)\\
    \hline
  	\end{tabular}
  	\caption[Basic parameters of the $Q_{\rm weak}$ experiment]{Basic parameters and typical operating conditions of the $Q_{\rm weak}$ experiment~\cite{qweak_proposal_2007}. The nominal beam current during production data taking was 180 $\mu$A, but for this commissioning dataset the current was 145 $\mu$A.}
  \label{tab:qweak_parameters}
\end{center}
\end{table}

\section{Analysis}
\label{Analysis}

The measured asymmetry ($A_{msr}$) can be derived (eq.~\ref{equ:eqAsym4}) from the raw asymmetry $A_{raw}$, which was calculated over each helicity quartet from the PMT integrated charge and normalized to beam charge $Y_{\pm}$. $A_{T}$ is the remnant transverse asymmetry and is separately measured using transversely polarized beam. Here, $A_{L}$ accounts for potential nonlinearity in the PMT response. 
The $ \Delta\chi_{i}$ are the helicity correlated differences in beam trajectory parameters (X, X$^{\prime}$, Y, Y$^{\prime}$) or energy (E) over the helicity quartet. The slopes $\frac{\partial A}{\partial\chi_{i}}$ were determined from linear regression using the natural and driven beam motion \cite{refNuruzzamanBmod} of the beam. The regression corrections $A_{reg}$ were obtained using differences and slopes and applied at the helicity quartet level. 

\begin{equation} \label{equ:eqAsym4}
\begin{split}
A_{msr} = \frac{Y_{+} - Y_{-}}{Y_{+} + Y_{-}} + A_{T} + A_{L} + \displaystyle\sum_{i=1}^{5} \left( \frac{\partial A}{\partial\chi_{i}}\right) \Delta\chi_{i}\\
\end{split}
\end{equation}

\begin{equation} \label{equ:eqAsym5}
A_{msr} = A_{raw} + A_{T} + A_{L} + A_{reg}
\end{equation}


\begin{figure}[!h]
\centering
\includegraphics[width=14.0cm]{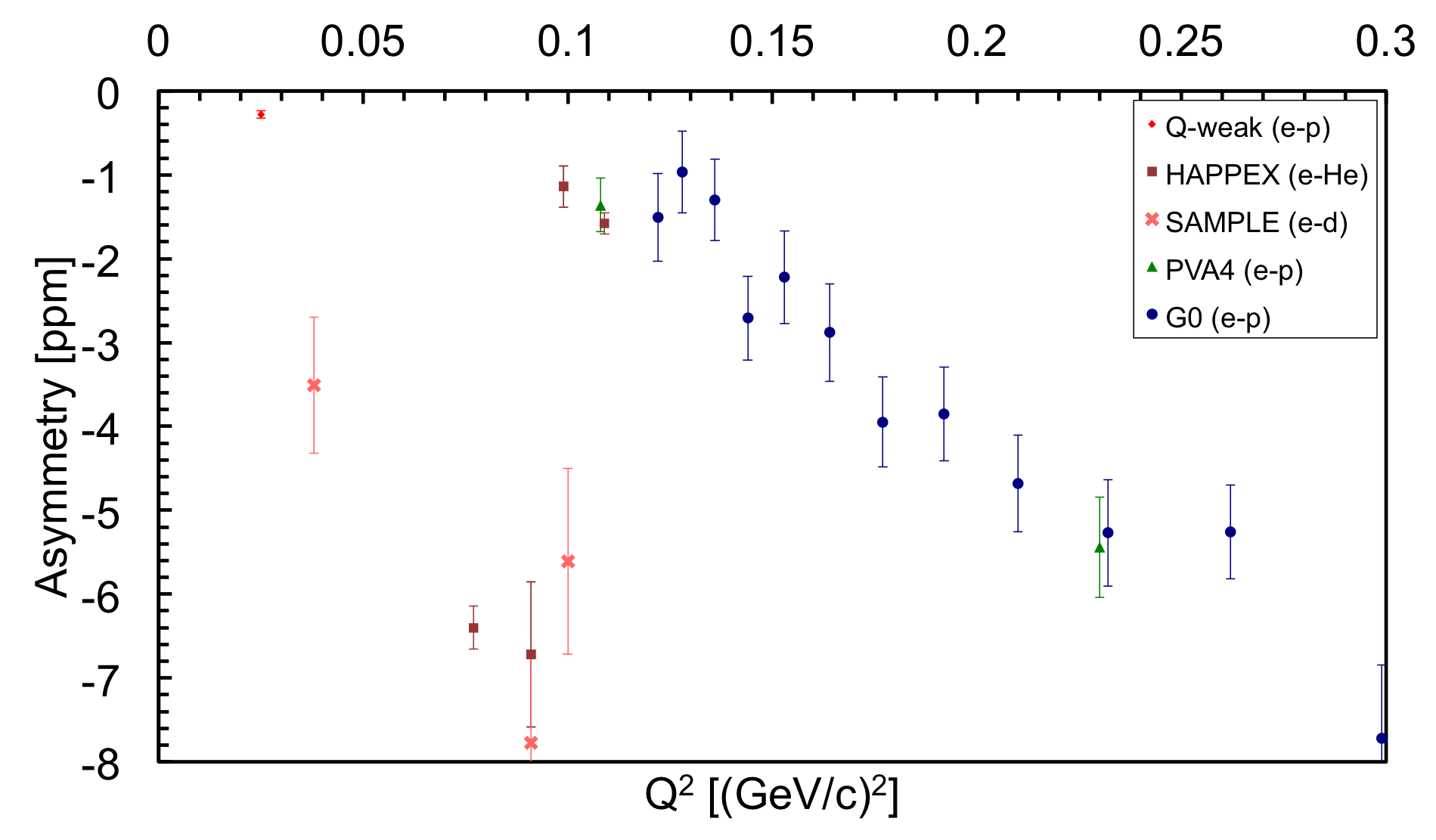}
\caption{(Color online) Asymmetries obtained from the commissioning dataset of the $Q_{\rm weak}$ and other PVES experiments up to $Q^{2}$ = 0.63 $(GeV/c)^{2}$ \cite{refSAMPLE1,refSAMPLE2,refHAPPEX1,refHAPPEX2,refHAPPEX3,refHAPPEX4,refHAPPEX5,refG01,refG02,refPVA41,refPVA42,refPVA43} were used to extract $Q_{W}^{p}$. A zoomed asymmetry \textit{vs} $Q^{2}$ is shown here for better visualization. The measured absolute asymmetry and absolute uncertainty from this measurement are $\sim$3 times smaller than the previous PVES measurements. }
\label{figQweakAsymmetry}
\end{figure}

\begin{equation} \label{equ:eqAsym6}
A_{ep} = R_{total} \left[  \frac{A_{msr}/P - \displaystyle\sum_{i=1}^{4} f_{bi}A_{bi} }{1- \displaystyle\sum_{i=1}^{4} f_{bi} } \right] 
\end{equation}


The PV asymmetry can be extracted from the measured asymmetry after correcting for the beam polarization, false asymmetries and backgrounds as shown in eq.~\ref{equ:eqAsym6}. The beam polarization was $\sim$89\% and was measured using a M{\footnotesize$\varnothing$}ller polarimeter~\cite{refMoller}. 
Here the multiplicative correction $R_{total}$ = $R_{RC}R_{Det}R_{Bin}R_{Q^{2}}$. $R_{RC}$ = 1.010 $\pm$ 0.005 is a radiative correction deduced from simulations with and without bremsstrahlung as described in Refs. \cite{refHAPPEX1,refHAPPEX6}. $R_{Det}$ = 0.987 $\pm$ 0.007 accounts for the measured light variation and non-uniform $Q^{2}$ distribution across the detector bars. $R_{Bin}$ = 0.980 $\pm$ 0.010 is an effective kinematics correction \cite{refHAPPEX6} that corrects the asymmetry from <A($Q^{2}$)> to A(<$Q^{2}$>), and $R_{Q^{2}}$ = 1.000 $\pm$ 0.030 represents the precision in calibrating the central $Q^{2}$. The background corrections were coming from target windows (b1), beamline scattering (b2), other neutral background (b3), and inelastics (b4). The contribution of each background to the measured asymmetry is shown in Table~\ref{tab-3}. 
The most significant correction for this dataset was from the aluminum target windows.

\begin{figure}
\centering
\includegraphics[width=14.0cm]{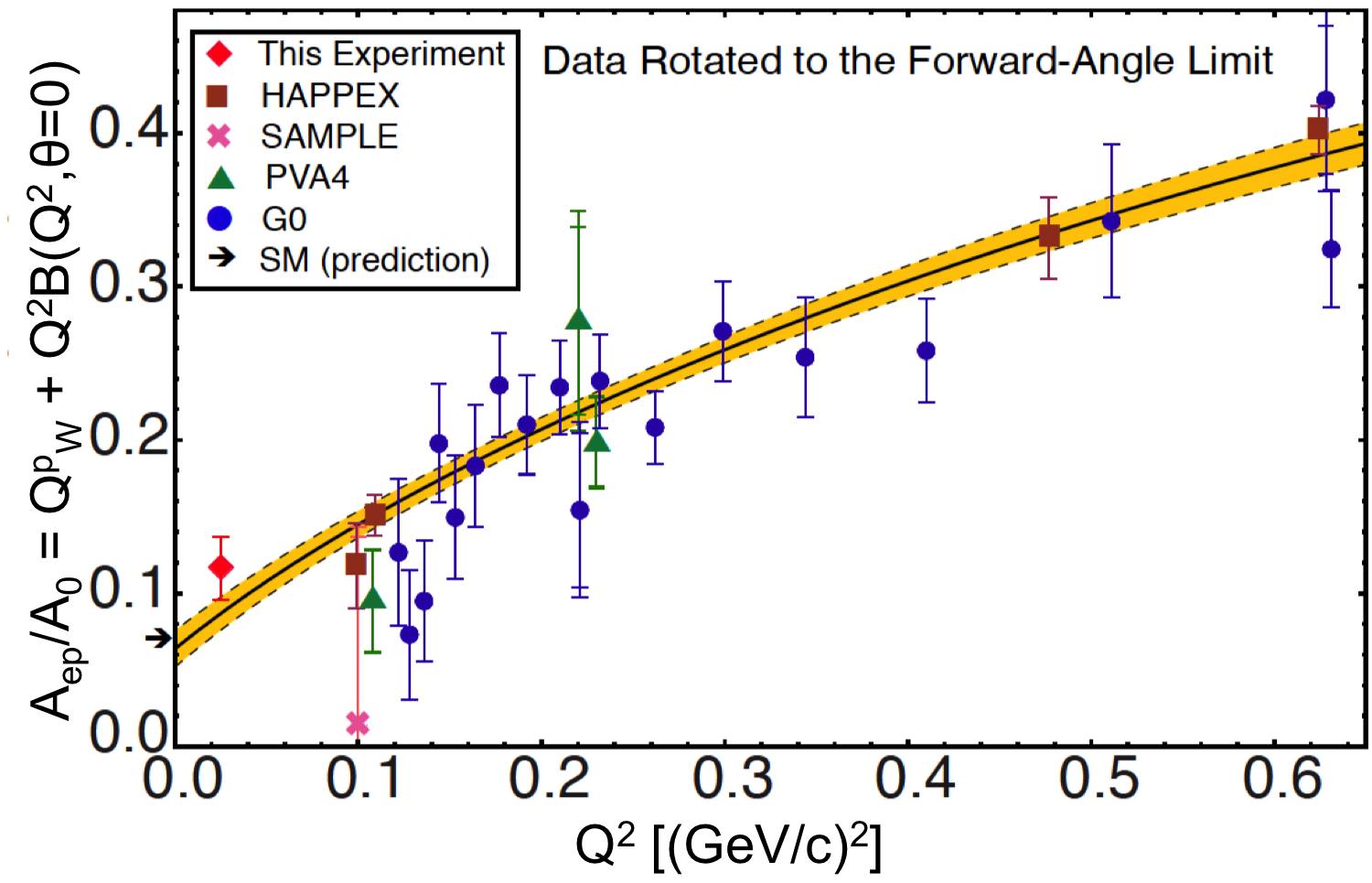}
\caption{(Color online) Global fit result (solid line) presented in the forward angle limit as reduced asymmetries derived from this measurement as well as other PVES experiments up to $Q^{2}$ = 0.63 $(GeV/c)^{2}$ \cite{refSAMPLE1,refSAMPLE2,refHAPPEX1,refHAPPEX2,refHAPPEX3,refHAPPEX4,refHAPPEX5,refG01,refG02,refPVA41,refPVA42,refPVA43}, including proton, helium, and deuterium data. The additional uncertainty arising from this rotation is indicated by outer error bars on data points. The band indicates the uncertainty in the fit. $Q^{p}_{W}$ is the intercept of the fit. The SM prediction \cite{PDG2012} is also shown (arrow).}
\label{figQweakResult25Percent}
\end{figure}


\begin{SCfigure}
\centering
\includegraphics[width=8.0cm]{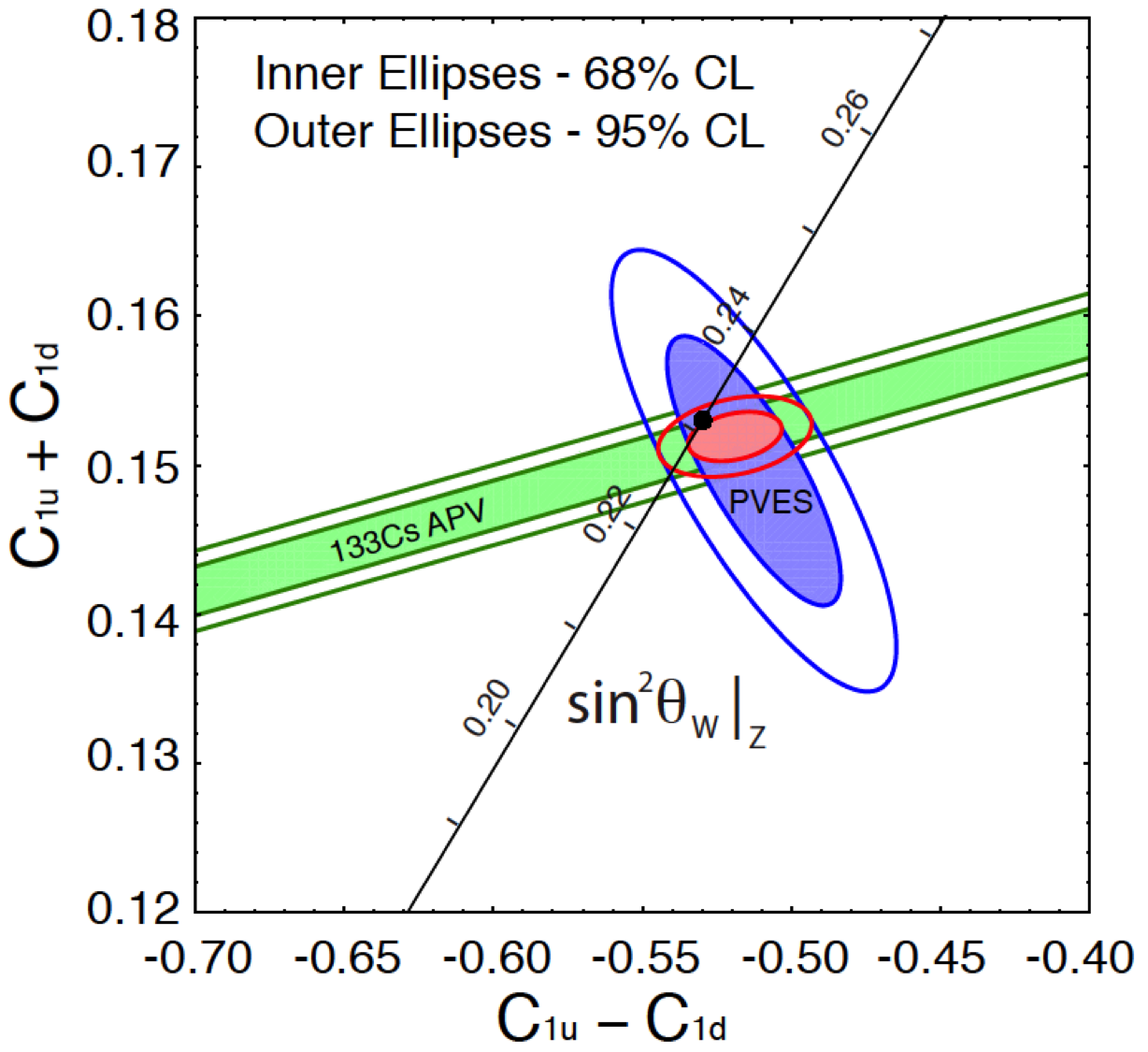}
\caption{
(Color online) Constraints on the neutral-weak quark coupling constants {$C_{1u}- C_{1d}$} (isovector) and 
{$C_{1u} + C_{1d}$} (isoscalar). 
The near horizontal (green) APV band constrains the isoscalar combination from $^{133}$Cs data~\cite{refCsData,refCsData2}.
The vertical (blue) ellipse 
represents the global fit of the existing $Q^2 \textless 0.63$ $(GeV/c)^{2}$ PVES data~\cite{refSAMPLE1,refSAMPLE2,refHAPPEX1,refHAPPEX2,refHAPPEX3,refHAPPEX4,refHAPPEX5,refG01,refG02,refPVA41,refPVA42,refPVA43} including the new result reported here at $Q^{2}$=0.025 $(GeV/c)^{2}$.
The small (red) ellipse near the center of the figure shows the result obtained by combining the APV and PVES information. 
The SM prediction~\cite{PDG2012} as a function of $\sin^{2}\theta_W $ in the $\overline{MS}$
scheme is plotted (diagonal black line) with the SM best fit value indicated by the (black) point at $\sin^{2}\theta_W $=0.23116. 
}
\label{figQweakQuarkCouplings}
\end{SCfigure}


\begin{table}[!h]
\centering
\begin{tabular}{ l | c  c  r }
\hline
 & Correction & \multicolumn{2}{c}{Contribution} \\
  & Value [ppb] & \multicolumn{2}{c}{to $\Delta A_{ep}$ [ppb]} \\ 
  \hline 
  \multicolumn{4}{c}{Normalization Factors Applied to $A_{raw}$} \\ 
  \hline 
Beam Polarization 1/P & -21 & \multicolumn{2}{r}{5} \\
Kinematics $R_{total}$ & 5 & \multicolumn{2}{r}{9} \\ 
Background Dilution 1/(1-$f_{total}$) & -7 & \multicolumn{2}{r}{-} \\ 
\hline
\multicolumn{4}{c}{Asymmetry Corrections} \\
\hline  
Beam Asymmetries $\kappa A_{reg}$ & -40 & \multicolumn{2}{r}{13} \\ 
Transverse Polarization $\kappa A_{T}$ & 0 & \multicolumn{2}{r}{5} \\ 
Detector Linearity $\kappa A_{L}$ & 0 & \multicolumn{2}{r}{4} \\ 
\hline
Backgrounds  & $\kappa Pf_{bi}A_{bi}$ & $\delta(f_{bi})$ & $\delta(A_{bi})$ \\ 
\hline 
\hspace{0.1cm} Target Windows ($b_{1}$) & -58 & 4 & 8 \\ 
\hspace{0.1cm} Beamline Scattering ($b_{2}$) & 11 & 3 & 23 \\ 
\hspace{0.1cm} Other Neutral Bkg ($b_{3}$) & 0 & 1 & \textless 1 \\ 
\hspace{0.1cm} Inelastics ($b_{4}$) & 1 & 1 & \textless 1 \\ 
\hline

\end{tabular}
\caption{Summary of corrections and the associated systematic uncertainty in ppb for the commissioning dataset. The table shows the contributions of normalization factors on $A_{raw}$. The properly normalized contributions from regression, transverse leakage, and detector nonlinearity are shown in the middle section. The background correction terms listed here include only $R_{tot} f_{i}A_{i}/(1 - f_{tot})$. The uncertainties in $A_{ep}$ due to dilution fraction and background asymmetry uncertainties are noted separately~\cite{qweak25percent}.}
\label{tab-3}       
\end{table}

\section{Result}
\label{Result}

The measured parity violating asymmetry, $A_{ep}$, using the commissioning data (eq.~\ref{equ:eqAsym6}) at <$Q^{2}$> = 0.0250 $\pm$ 0.0006 $(GeV/c)^{2}$ is -279 $\pm$ 35 (statistics) $\pm$ 31 (systematics) parts per billion (ppb). 
This is the most precise absolute parity violating electron-proton scattering measurement to date (Figure~\ref{figQweakAsymmetry}).
This reported asymmetry was then combined with other 
PVES results \cite{refSAMPLE1,refSAMPLE2,refHAPPEX1,refHAPPEX2,refHAPPEX3,refHAPPEX4,refHAPPEX5,refG01,refG02,refPVA41,refPVA42,refPVA43} on hydrogen, deuterium, and helium in a global fit following the prescription in [4] to extract the weak charge of the proton using eq.~\ref{equ:eqAsym3} and is shown in Figure~\ref{figQweakResult25Percent}. 
The angle dependence of the strange and axial form-factor contributions was removed by subtracting [$A_{calc}$($\theta$, $Q^{2}$) - $A_{calc}$(0, $Q^{2}$)] from $A_{ep}$($\theta$, $Q^{2}$) to demonstrate the two-dimensional global fit ($\theta$, $Q^{2}$) in a single dimension ($Q^{2}$). The calculated asymmetries $A_{calc}$ are determined from eq.~\ref{equ:eqAsym2} using the results of the fit.
The extracted weak charge of the proton using the measured asymmetry from this experiment and PVES data is $Q^{p}_{W}$ = 0.064 $\pm$ 0.012~\cite{qweak25percent}. At the effective kinematics the predicted weak charge by the SM \cite{PDG2012} $Q^{p}_{W}$(SM) is 0.0710 $\pm$ 0.0007 as shown by an arrow in Figure~\ref{figQweakResult25Percent}.

The present measurement also constrains the neutral weak quark couplings. The result of a fit combining the most recent correction~\cite{refCsData} to the $^{133}$Cs APV result~\cite{refCsData2} with the world PVES data (including the present measurement), is shown in Figure~\ref{figQweakQuarkCouplings}.
The neutral-weak couplings determined from this combined fit are $C_{1u}$ = 0.1835 $\pm$ 0.0054 and $C_{1d}$ = 0.3355  $\pm$ 0.0050, with a correlation coefficient 0.980. In addition, combining the $C_{1}$'s the neutron's weak charge was extracted as $Q_{W}^{n}$(PVES+APV) =  -2($C_{1u}$ + 2$C_{1d}$) = - 0.9750 $\pm$0.010 which is consistent with $Q_{W}^{n}$(SM) = - 0.9890$\pm$0.0007~\cite{qweak25percent}.
The commissioning results reported here are derived from only about 4\% of the data that were collected for the experiment. The expected final asymmetry is anticipated to have much smaller uncertainty with full statistics and improved analysis of systematic uncertainties. 
These data will also be used to determine the weak-mixing angle, $\sin^{2} \theta_{W}$, with a relative uncertainty of $\textless$0.5\% that will provide a competitive measurement of the running of $\sin^{2} \theta_{W}$ to low $Q^{2}$.

This work was supported by DOE contract No. DE-AC05-06OR23177, under which Jefferson Science Associates, LLC, operates Thomas Jefferson National Accelerator Facility. I would like to thank the $Q_{\rm weak}$ collaboration for the opportunity to present on its behalf.

\end{document}